\documentclass[english,final,table]{IEEEtran}
\usepackage[T1]{fontenc}
\usepackage[latin9]{inputenc}
\usepackage{xcolor}
\usepackage{pdfcolmk}
\usepackage{array}
\usepackage{amsmath}
\usepackage{amsthm}
\usepackage{graphicx}
\PassOptionsToPackage{normalem}{ulem}
\usepackage{ulem}

\makeatletter

\providecolor{lyxadded}{rgb}{0,0,1}
\providecolor{lyxdeleted}{rgb}{1,0,0}

\DeclareRobustCommand{\lyxdeleted}[3]{{\color{lyxdeleted}\lyxsout{#3}}}
\DeclareRobustCommand{\lyxsout}[1]{\ifx\\#1\else\sout{#1}\fi}

\usepackage{amsmath}
\usepackage{amssymb}
\usepackage[level=0]{wgroup_message}  
\usepackage[table]{xcolor}
\definecolor{lightcyan}{rgb}{0.88, 1.0, 1.0}
\usepackage{booktabs}

\author{\IEEEauthorblockN{Junting Chen, Bowen Li, Hao Sun,
Shuguang Cui, Nikolaos Pappas}
\thanks{This work was supported in part by the National Natural Science Foundation of China under Grant 62293482 and Grant 62171398, in part by Guangdong Basic and Applied Basic Research Foundation under Grant 2024A1515011206, in part by Shenzhen Science and Technology Program under Grant KJZD20230923115104009, in part by Guangdong Provincial Key Laboratory of Future Networks of Intelligence under Grant 2022B1212010001, in part by Shenzhen Key Laboratory of Big Data and Artificial Intelligence under Grant SYSPG20241211173853027, in part by Guangdong Province Radio Science Data Center, in part by the ELLIIT, and in part by the European Union (6G-LEADER) under Grant 101192080. (Corresponding author: Bowen Li.)}
\thanks{Junting Chen and Shuguang Cui are with the School of Science and Engineering (SSE), 
and the Shenzhen Future Network of Intelligence Institute (FNii--Shenzhen), The Chinese University of Hong Kong (Shenzhen), Shenzhen, Guangdong 518172, China.
(email: juntingc@cuhk.edu.cn, shuguangcui@cuhk.edu.cn).}
\thanks{Bowen Li and Nikolaos Pappas are with Department of Computer and Information Science, Link{\"o}ping University, 58183, Link{\"o}ping, Sweden. (email: bowen.li@liu.se, nikolaos.pappas@liu.se).}
\thanks{Hao Sun is with the Department of Electrical Engineering, City University of Hong Kong, Hong Kong. (email: hao.sun@cityu.edu.hk).}
}


\usepackage[acronym]{glossaries}
\newcommand{\newac}{\newacronym}
\newcommand{\ac}{\gls}

\newcommand{\acpl}{\glspl}

\newac{speb}{SPEB}{square position error bound}
\newac[plural=EFIMs,firstplural=Fisher information matrices (EFIMs)]{efim}{EFIM}{Fisher information matrix}
\newac{ne}{NE}{Nash equilibrium}
\newac{mse}{MSE}{mean squared error}
\newac{toa}{TOA}{time-of-arrival}
\newac{snr}{SNR}{signal-to-noise ratio}
\newac{lan}{LAN}{local area network}
\newac{psd}{PSD}{positive semidefinite}
\newac{pd}{PD}{positive definite}
\newac{wrt}{w.r.t.}{with respect to}
\newac{lhs}{L.H.S.}{left hand side}
\newac{wp1}{w.p.1}{with probability 1}
\newac{kkt}{KKT}{Karush-Kuhn-Tucker}
\newac{wlog}{w.l.o.g.}{without loss of generality}
\newac{mle}{MLE}{maximum likelihood estimation}
\newac{gps}{GPS}{global positioning system}
\newac{rssi}{RSSI}{received signal strength indicator}
\newac{mimo}{MIMO}{multiple-input multiple-output}
\newac{csi}{CSI}{channel state information}
\newac{fdd}{FDD}{frequency division duplexing}
\newac{ms}{MS}{mobile station}
\newac{bs}{BS}{base station}
\newac{d2d}{D2D}{device-to-device}
\newac{slnr}{SLNR}{signal-to-interference-leakage-and-noise-ratio}
\newac{ula}{ULA}{uniform linear antenna array}
\newac{pas}{PAS}{power angular spectrum}
\newac{mmse}{MMSE}{minimum mean square error}
\newac{zf}{ZF}{zero-forcing}
\newac{rzf}{RZF}{regularized zero-forcing}
\newac{as}{AS}{angular spread}
\newac{aod}{AOD}{angle of departure}
\newac{iid}{i.i.d.}{independent and identically distributed} 
\newac{sinr}{SINR}{signal-to-interference-and-noise ratio}
\newac{tdd}{TDD}{time-division duplex}
\newac{rvq}{RVQ}{random vector quantization}
\newac{rhs}{R.H.S.}{right hand side}
\newac{mrc}{MRC}{maximum ratio combining}
\newac{cdf}{CDF}{cumulative distribution function}
\newac{a.s.}{a.s.}{almost surely}
\newac{los}{LOS}{line-of-sight}
\newac{jsdm}{JSDM}{joint spatial division and multiplexing}
\newac{map}{MAP}{maximum a posteriori}
\newac{klt}{KLT}{Karhunen-Lo\`eve Transform}
\newac{lbe}{LBE}{link bargaining equilibrium}
\newac{se}{SE}{Stackelberg equilibrium}
\newac{uav}{UAV}{unmanned aerial vehicle}
\newac{nlos}{NLOS}{non-line-of-sight}
\newac{pdf}{PDF}{probability density function}
\newac{em}{EM}{expectation-maximization}
\newac{knn}{KNN}{$k$-nearest neighbor}
\newac{svd}{SVD}{singular value decomposition}
\newac{nmf}{NMF}{non-negative matrix factorization}
\newac{umf}{UMF}{unimodality-constrained matrix factorization}
\newac{rmse}{RMSE}{rooted mean squared error}
\newac{olos}{OLOS}{obstructed line-of-sight}
\newac{mmw}{mmW}{millimeter wave}
\newac{ber}{BER}{bit error rate}
\newac{rss}{RSS}{received signal strength}
\newac{lp}{LP}{linear program}
\newac{ufw}{U-FW}{unimodal Frank-Wolfe}
\newac{utf}{UTF}{unimodality-constrained tensor factorization}
\newac{fw}{FW}{Frank-Wolfe}
\newac{iot}{IoT}{Internet-of-Things}
\newac{mae}{MAE}{mean absolute error}
\newac{crb}{CRB}{Cram\'er-Rao bound}
\newac{aoa}{AoA}{angle of arrival}
\newac{wcl}{WCL}{weighted centroid localization}

\setkeys{Gin}{width=1.0\columnwidth}

\renewcommand{\lyxdeleted}[3]{{\color{lyxdeleted}{}}}

\newac{mdp}{MDP}{Markov decision process}
\newac{wsn}{WSN}{wireless sensor network}
\newac{iff}{iff}{if and only if}
\newac{qsi}{QSI}{queue state information}
\newac{ue}{UE}{user equipment}
\newac{gbs}{GBS}{ground base station}
\newac{abs}{ABS}{aerial base station}
\newac{leo}{LEO}{low earth orbit}
\newac{gcs}{GCS}{ground control station}
\newac{tdma}{TDMA}{time division multiple access}
\newac{son}{SON}{self-organized network}
\newac{ofdm}{OFDM}{orthogonal frequency division multiplexing}
\newac{uma}{UMa}{urban macro}
\newac{isac}{ISAC}{integrated sensing and communication}
\newac{sca}{SCA}{successive convex approximation}
\newac{umi}{UMi}{Urban Micro}
\newac{lr}{LR}{Lagrangian relaxation}
\newac{qam}{QAM}{quadrature amplitude modulation}
\newac{tsp}{TSP}{travelling salesman problem}
\newac{qos}{QoS}{quality of service}
\newac{np}{NP}{non-deterministic polynomial-time}
\newac{bcd}{BCD}{block coordinate descent}
\newac{pss}{PSS}{primary synchronization signal}
\newac{sss}{SSS}{secondary synchronization signal}
\newac{cfo}{CFO}{carrier frequency offset}
\newac{ifft}{IFFT}{inverse fast fourier transform}
\newac{cp}{CP}{cyclic prefix}
\newac{nr}{NR}{new radio}
\newac{gmlr}{GMLR}{generalized maximum likelihood ratio}
\newac{ssb}{SSB}{synchronization signal block}
\newac{pci}{PCI}{physical cell identity}
\newac{nid2}{NID2}{secondary network identifier}
\newac{nid1}{NID1}{first network identifier}
\newac{ml}{ML}{maximum likelihood}
\newac{ls}{LS}{least square}
\newac{ap}{AP}{access point}
\newac{awgn}{AWGN}{additive white Gaussian noise}
\newac{ris}{RIS}{reconfigurable intelligent surface}
\newac{usrp}{USRP}{universal software radio peripheral}
\newac{sic}{SIC}{successive interference cancellation}
\newac{crlb}{CRLB}{Cram\'{e}r-Rao lower bound}
\newac{fim}{FIM}{Fisher information martix}
\newac{lae}{LAE}{low-altitude economy}
\newac{evtol}{eVTOL}{electric vertical takeoff and landing}
\newac{amc}{AMC}{adaptive modulation and coding}
\newac{rsma}{RSMA}{rate splitting multiple access}
\newac{marl}{MARL}{multi-agent reinforcement learning}
\newac{1d}{1D}{one-dimensional}
\newac{2d}{2D}{two-dimensional}
\newac{3d}{3D}{three-dimensional}
\newac{4d}{4D}{four-dimensional}
\newac{5d}{5D}{five-dimensional}
\newac{6d}{6D}{six-dimensional}
\newac{7d}{7D}{four-dimensional}
\newac{laan}{LAAN}{low altitude aircraft network}

\usepackage[noadjust]{cite}
\usepackage{color}  
\usepackage{graphicx,psfrag,cite,subfigure}

\definecolor{lightcyan}{rgb}{0.88, 1.0, 1.0}

\makeatother

\usepackage{babel}
\begin{document}
\title{Predictive Communications for \\Low-Altitude Networks}
\maketitle

\begin{abstract}
The emergence of dense, mission-driven aerial networks supporting the low-altitude economy presents unique communication and security challenges, including extreme channel dynamics and severe cross-tier interference. Traditional reactive communication paradigms are ill-suited to these environments, as they fail to leverage the network's inherent predictability. This paper introduces {\em predictive communication}, a novel paradigm transforming network management from reactive adaptation to proactive optimization. The approach is enabled by fusing predictable mission trajectories with stable, large-scale radio environment models ({\em e.g.}, radio maps). Specifically, we present a hierarchical framework that decomposes the predictive cross-layer resource allocation problem into three layers: strategic (routing), tactical (timing), and operational (power). This structure aligns decision-making timescales with the accuracy levels and ranges of available predictive information. We demonstrate that this foresight-driven framework achieves an order-of-magnitude reduction in cross-tier interference and enables proactive security against threats such as jamming and spoofing, laying the groundwork for efficient, resilient, and secure low-altitude communication systems.
\end{abstract}

\section{Introduction}

The emergence of the low-altitude economy, an ecosystem of services conducted by \acpl{uav} and \ac{evtol} aircraft, is set to redefine urban and rural infrastructure \cite{SonLinWanSun:M25,JiaLiZhuLi:M25}. From automated logistics to real-time infrastructure inspection, these services depend on the seamless operation of extensive low-altitude networks. As these networks form a dynamic aerial layer above the terrestrial systems \cite{VaeLinZha:J24,SonZenYanRen:J25}, {a communication paradigm that ensures efficient, reliable, and secure operation becomes paramount.}

\begin{figure*}
\begin{centering}
\includegraphics[width=0.95\textwidth]{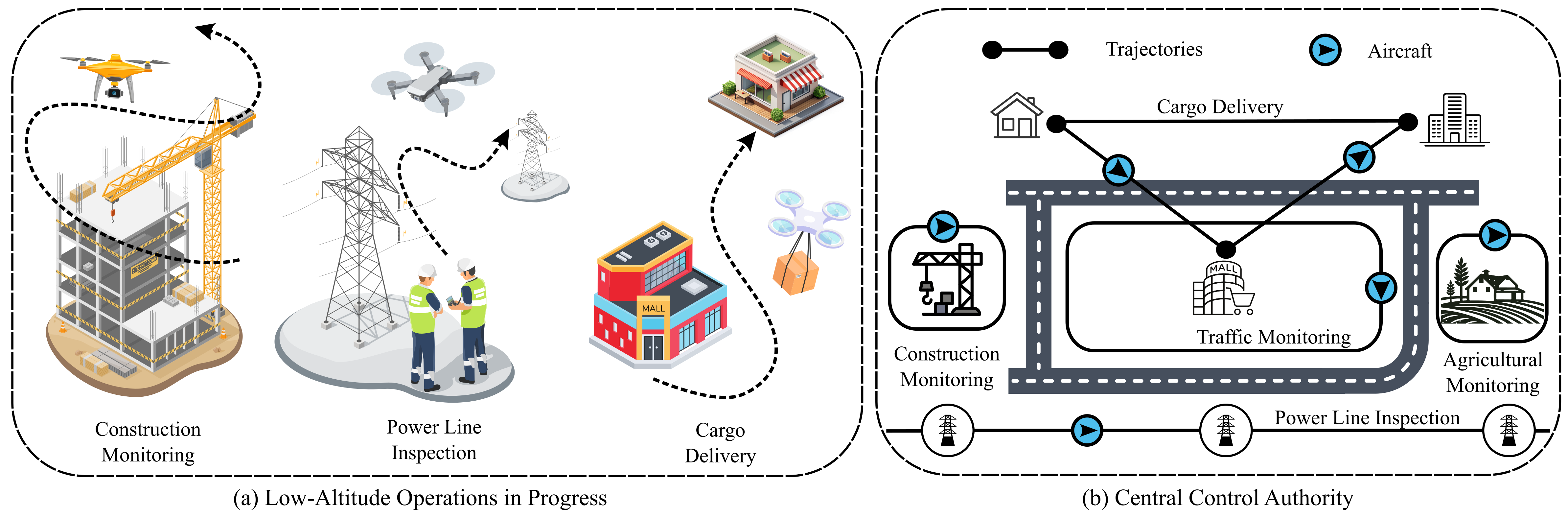}
\par\end{centering}
\caption{\label{fig:LAE_LAAN}The operational concept of the low-altitude network and its implications for predictive communication. (a) Diverse low-altitude operations, such as infrastructure inspection and cargo delivery, are performed by aircraft following predetermined trajectories that are tailored to their mission requirements. (b) For reasons of air traffic safety and operational management, these diverse, mission-driven flight plans are filed with a central control authority, which possesses a priori, network-wide knowledge of all aircraft movements. This centralized oversight of a predictable, time-varying network topology creates the foundational opportunity for predictive communication.}
\end{figure*}

\begin{figure*}
\begin{centering}
\includegraphics[width=0.95\textwidth]{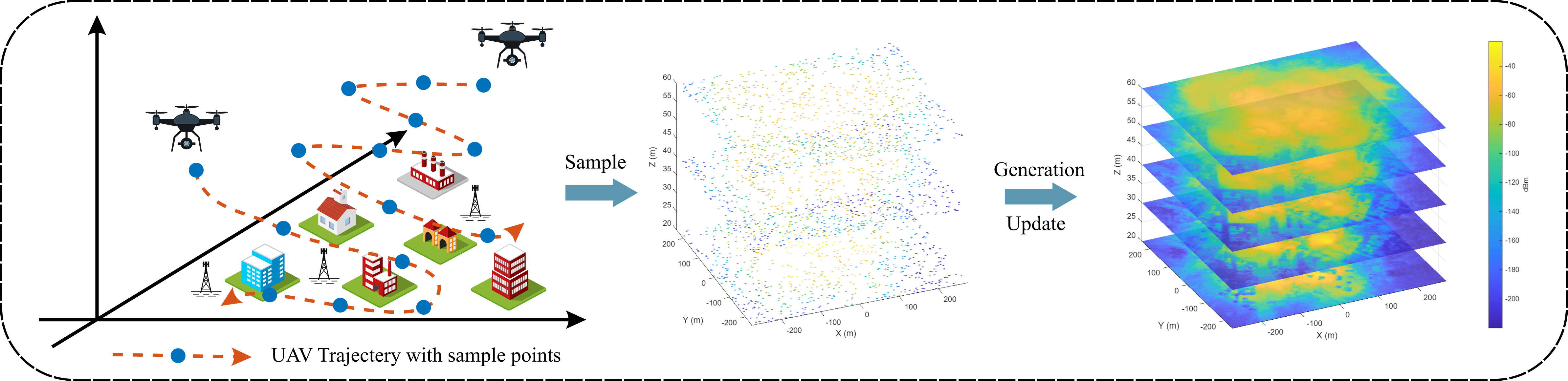}
\par\end{centering}
\caption{\label{fig:Radio_Map}Illustration of an aerial-assisted radio map generation and update process. Multiple \acpl{uav} collaborate to sample the radio environment along their trajectories, creating a sparse point cloud of channel data. This aggregated data is then used to generate or update a dense, multi-layered environmental model, enabling a continuously refined understanding of the radio environments.}
\end{figure*}

However, low-altitude networks present a unique hybrid challenge that is not addressed by existing paradigms. They are far more dynamic than conventional terrestrial networks \cite{SinKumMajSat:J25,HanJiaYu:J25}, with high-speed, 3D\footnote{Throughout this paper, we use the notation xD to denote an x-dimensional space.} mobility that induces drastic channel variations. However, they are more constrained and operate in a much more complex environment than satellite networks. While space systems \cite{HanXuZhaWan:J23, NtoTomRui:C24} benefit from near-perfect trajectory predictability in an electromagnetically sparse void, low-altitude aircraft face unpredictable wind, dynamic obstacles, and a dense, interference-coupled airspace that invalidates the assumptions of traditional satellite communication design.

The opportunity to master this environment lies in harnessing a powerful, yet underutilized, resource: {\em imperfect but statistically robust predictability}. This foresight is derived from two pillars: the a priori knowledge of mission-driven flight plans filed with a central authority \cite{LiChe:J24b,LiuXuWanChe:J20,LiChe:J24}, as illustrated in Fig.~\ref{fig:LAE_LAAN}, and the quasi-static nature of the large-scale radio environment, which can be captured in models like radio maps \cite{HeAiGuaWan:J19,SunChe:J24,WanZhaNieYu:M25}, as illustrated in Fig.~\ref{fig:Radio_Map}. This available foresight allows us to shift from a reactive to a proactive design philosophy fundamentally.

This paper introduces a layered {\em predictive communication} framework that is explicitly designed to harness this imperfect foresight. The key to unlocking unprecedented efficiency and control is to hierarchically decompose the network management problem in a way that matches the intrinsic structure of the predictive information itself. Our framework translates the coarse, long-horizon global predictions into strategic directives, while leveraging precise, short-horizon local data for tactical execution and operational efficiency. Our key contributions are made as follows:
\begin{itemize}
    \item {We formalize the concept of \emph{predictive communication} for low-altitude networks, establishing its feasibility upon two pillars: a priori mission trajectories and large-scale radio environment models. We also analyze the hierarchical nature of this predictive information, revealing a fundamental trade-off between predictive range and accuracy that governs system design.}
    \item {We propose a novel layered predictive communication framework that is structurally matched to this information hierarchy. The framework decomposes the complex network management problem into three distinct layers: strategic long-term routing, tactical mid-term coordination, and operational short-term policy optimization, enabling a robust and efficient use of imperfect foresight.}
    \item {We demonstrate our framework's dual utility. Its effectiveness is quantitatively validated in a case study on interference mitigation, while its versatility is conceptually demonstrated by outlining how the same architecture, through tailored objectives and constraints, can proactively address security threats like jamming, spoofing, and interception.}
\end{itemize}

{The rest of this article is organized as follows. We introduce the predictive information and its hierarchical nature. Then, the proposed framework and security-oriented extension are presented. Finally, we provide some future directions and conclude the article.}

\section{Foundations of Predictive Communication: Building Actionable Foresight}\label{Sec:Fundamental}

\begin{figure*}
\begin{centering}
\includegraphics[width=1\textwidth]{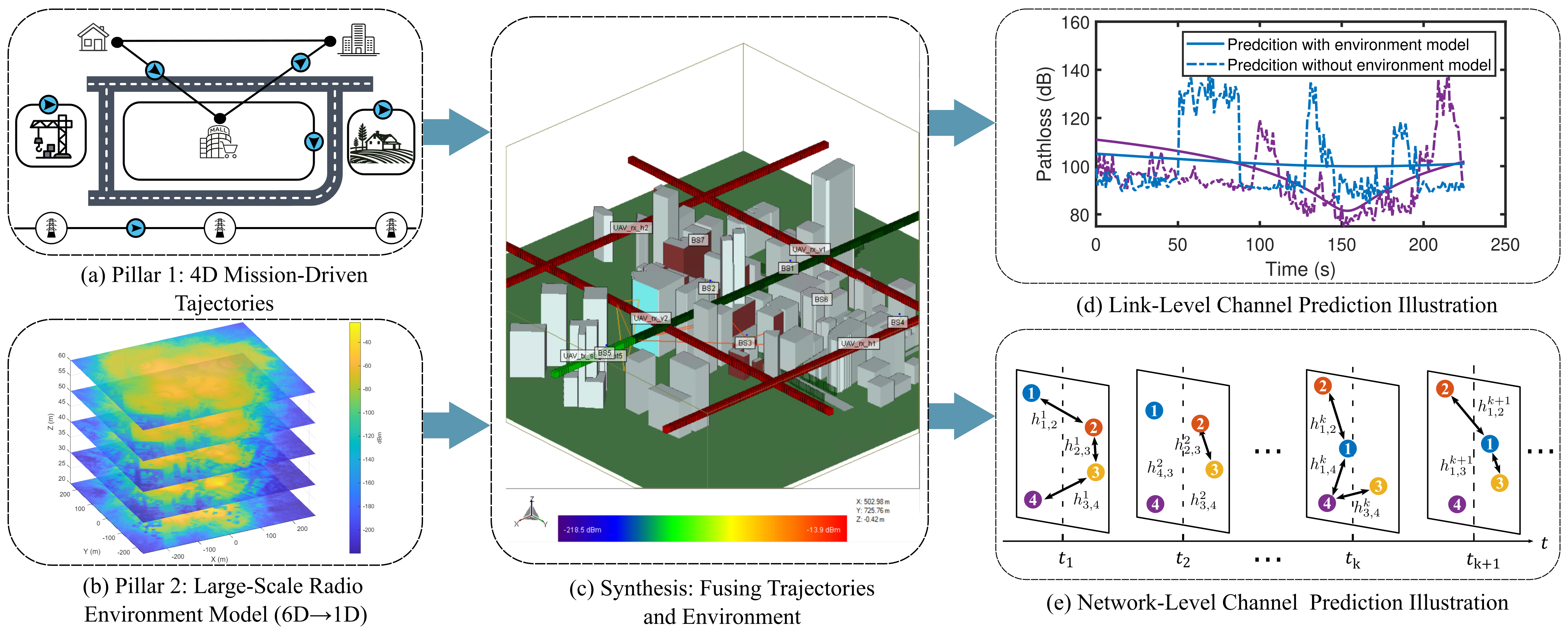}
\par\end{centering}
\caption{\label{fig:feasibility_predictive}Generation process of the predictive large-scale \ac{csi}, illustrating the information processing flow. The architecture leverages two foundational inputs: (a) a set of 4D mission trajectories (3D space + 1D time) for all aircraft, and (b) a 6D radio environment model that maps the 6D spatial coordinates of a transmitter-receiver pair to a 1D prediction of large-scale channel quality. In the (c) synthesis step, these inputs are fused. For any two aircraft, their independent 4D trajectories are first temporally aligned to form a 7D state descriptor (3D transmitter position, 3D receiver position, and 1D time). At each time step, the 6D spatial component is mapped by the environmental model to a 1D channel quality value. This process transforms the initial trajectory data into two representative predictive views: a link-level channel prediction, which shows the 2D (channel quality vs. time) forecast for a series of single links, and a network-level channel prediction, which displays the evolution of the entire low-altitude network topology.}
\end{figure*}

The technical feasibility of predictive communication relies on the ability to forecast future channel states. Two key pillars of predictability make this possible in low-altitude networks: the deterministic nature of mission-driven aircraft trajectories and the statistical stability of the large-scale radio environment. This section formally establishes these foundations. We will first introduce each of the two pillars and then detail the synthesis process that combines them to produce a predictive channel forecast, as illustrated in the workflow of Fig.~\ref{fig:feasibility_predictive}.

\subsection{Pillar 1: Trajectory Predictability}

The first pillar of our predictive communication architecture is the set of predetermined 4D trajectories, that is, three spatial coordinates plus time, for every low-altitude aircraft in the network. The availability and utility of this information are underpinned by two practical facts inherent to the low-altitude networks.
\begin{itemize}
    \item Mission-driven flight planning. Most low-altitude aircraft are deployed for clearly defined tasks, ensuring their flight paths are well-planned. These plans are filed with a central control authority for air-traffic safety and operational management reasons.
    \item Statistical adherence to the flight plan. Although wind, temporary restrictions, or ad-hoc directives may cause brief deviations, aircraft typically re-converge to their planned routes. Over the timescales relevant to resource planning, the realized trajectory aligns, on average, with the pre-filed path.
\end{itemize}

These principles are illustrated in Fig.~\ref{fig:LAE_LAAN}. Various mission-driven tasks are depicted in Fig.~\ref{fig:LAE_LAAN}(a), such as cargo aircraft following fixed logistics routes and inspection aircraft conducting repetitive scans over infrastructure. The architectural implication of this operational model is shown in Fig.~\ref{fig:LAE_LAAN}(b), where all flight plans in low-altitude networks are filed with and orchestrated by a central control authority. This entity, therefore, possesses network-wide, a priori knowledge of all planned aircraft movements. Owing to the high statistical adherence to these plans, the conceptual architecture in Fig.~\ref{fig:LAE_LAAN}(b) can be regarded as both a static plan and a snapshot of the network's expected real-time evolution.

Taken together, these two facts allow us to treat the complete set of 4D trajectories as a deterministic and statistically robust forecast of future node locations. In our architecture, this information forms the time-evolving skeleton of the network topology. It serves as a primary input, as illustrated in Fig.~\ref{fig:feasibility_predictive}(a), upon which all subsequent channel predictions are built.

\subsection{Pillar 2: Radio Environment Predictability}

The second pillar of our architecture is a predictable model of the radio environment that maps every transmitter-receiver coordinate pair to a statistical description of the large-scale channel. Formally, it is a static function from a 6D spatial input, the 3D coordinates for the transmitter and the receiver, to \ac{csi} statistics, including path loss, shadowing, {\em etc}. Two observations justify this pillar.
\begin{itemize}
    \item Quasi-static large-scale channel characteristics. Over minutes or hours, path-loss and shadowing between two fixed points remain effectively constant, even though small-scale fading fluctuates rapidly.
    \item Advanced radio-environment models now provide fine-grained, city-scale accuracy. State-of-the-art radio maps generated via ray tracing, measurement-driven interpolation, or high-fidelity digital twins can deliver metre-level resolution across the 3D workspace \cite{HeAiGuaWan:J19,SunChe:J24,WanZhaNieYu:M25}.
\end{itemize}

An aerial-assisted radio map is a powerful example of such a model, and its generation and update processes are illustrated in Fig.~\ref{fig:Radio_Map}. In this paradigm, aircraft act as mobile sensing agents, collecting geo-tagged channel measurements along their trajectories. This sparse dataset is collected and then transformed into a dense, multi-layered map at a central or edge repository using advanced radio-environment models that infer channel characteristics in unsampled locations. Crucially, this mechanism enables the continuous real-time update of the map, allowing it to adapt to environmental changes and reflect the most current state of the radio landscape.

Taken together, these facts allow us to treat the radio environment as a predictable, static information layer. As illustrated in Fig.~\ref{fig:feasibility_predictive}(b), this enables a central controller to query a database for the expected channel quality between any two predicted future locations, providing the second crucial input for our predictive synthesis.

\begin{table*}[t]
\caption{The Hierarchical Nature of Predictive Information
and Corresponding Proactive Tasks.}
\label{tab:h_nature}
\centering
\begin{tabular}{
  >{\centering\arraybackslash}p{0.16\textwidth}
  >{\centering\arraybackslash}p{0.14\textwidth}
  >{\centering\arraybackslash}p{0.23\textwidth}
  >{\centering\arraybackslash}p{0.17\textwidth}
  >{\centering\arraybackslash}p{0.17\textwidth}
}
\toprule
\textbf{Network Echelon} &
\textbf{Predictive Range} &
\textbf{Dominant Sources of Uncertainty} &
\textbf{Suitable Task Class} &
\textbf{Example Proactive Task} \\
\midrule
\rowcolor{lightcyan}
Central (e.g., Central Control Authority) &
Long-Term, Network-Level &
Trajectory Deviations, Environmental Modeling Errors, Channel Randomness &
Strategic, Network-Wide Planning &
End-to-End Routing \\
Local (e.g., Base Station, Cluster Head) &
Middle-Term, Group-Level &
Environmental Modeling Errors, Channel Randomness &
Tactical, Multi-Node Coordination &
Handover Timing \\
\rowcolor{lightcyan}
Individual (e.g., Aircraft) &
Short-Term, Link-Level &
Channel Randomness (after measurement) &
Link-Level Policy Optimization &
Power/Rate Control \\
\bottomrule
\end{tabular}
\end{table*}

\subsection{Synthesis: Fusing  Trajectories and Environment Model}

The synthesis of the two architectural pillars is a computational process that transforms deterministic knowledge about geometry and movement into actionable foresight about future communication quality. As depicted in Fig.~\ref{fig:feasibility_predictive}(c), this is achieved by systematically evaluating the static radio environment model (Pillar 2) at the future coordinates dictated by the time-varying mission trajectories (Pillar 1). This fusion effectively maps the dynamic network skeleton with predicted channel quality values, creating a unified, time-varying forecast of the network's \ac{csi}.

The output of this synthesis provides two complementary and essential predictive views. At a granular level, it yields link-level channel forecasts, which are time-series predictions of the quality for any specific link, as illustrated in the plot in Fig.~\ref{fig:feasibility_predictive}(d). At a holistic level, the aggregation of all such forecasts produces the final architectural output: the predictive, spatio-temporal channel graph (Fig.~\ref{fig:feasibility_predictive}(e)). This is achieved by mapping the predicted channel quality between every pair of nodes at each time instant to the weight of the corresponding edge in a network graph for that moment. This rich, 4D data structure represents the complete, time-indexed evolution of the network's connectivity and link qualities.

The successful generation of this graph is not merely a technical step; it is the proof of concept that establishes the fundamental feasibility of the predictive paradigm.

\section{Hierarchical Nature of Predictive Information}

The predictive architecture outlined in Section \ref{Sec:Fundamental} provides an idealized glimpse of future connectivity; however, real-world forecasts are inherently probabilistic. Factors such as modeling errors, trajectory deviations, and feedback delays introduce uncertainty that grows with the prediction horizon. Moreover, the information available to different network entities, from a central controller to an individual aircraft, varies in scope, freshness, and fidelity, yielding a hierarchical information structure governed by a fundamental range-accuracy trade-off. This section identifies the principal sources of uncertainty, presents a three-tier hierarchy of predictive entities, and shows how their distinct sensing and processing capabilities lead to the trade-off summarized in Table~\ref{tab:h_nature}.

\subsection{Fundamental Sources of Predictive Uncertainty}

The gap between an idealized forecast and real-world performance stems from three primary sources of uncertainty, each degrading predictive quality in a different way:
\begin{itemize}
\item Trajectory deviations: Although missions are pre-planned, an aircraft's realized path can drift from its filed trajectory due to wind gusts, temporary air-traffic directives, minor navigation corrections, or proactive security maneuvers, such as detouring around a newly identified high-risk area.
Even small lateral shifts are critical in the low altitude networks: a displacement of only a few meters may flip a link from \ac{los} to \ac{nlos}, causing significant errors in large-scale channel prediction.
\item Environmental model inaccuracy and staleness: State-of-the-art radio maps and digital twins provide high-fidelity, large-scale channel estimates, yet they are still abstractions of reality. In addition, the global model held by a central controller is updated only as frequently as it receives, often compressed, reports from aircraft. Consequently, its view is less fresh and less accurate than the locally refined model maintained on board each platform.
\item Channel randomness: Even with perfect location knowledge and a flawless environment model, wireless channels exhibit unpredictable small-scale fading induced by multipath propagation. This represents a fundamental physical limit: fine-grained channel states cannot be known without instantaneous measurement.
\end{itemize}

Together, these factors transform deterministic foresight into probabilistic insight, motivating the layered optimization framework developed in the sequel.

\begin{figure*}
\begin{centering}
\includegraphics[width=1\textwidth]{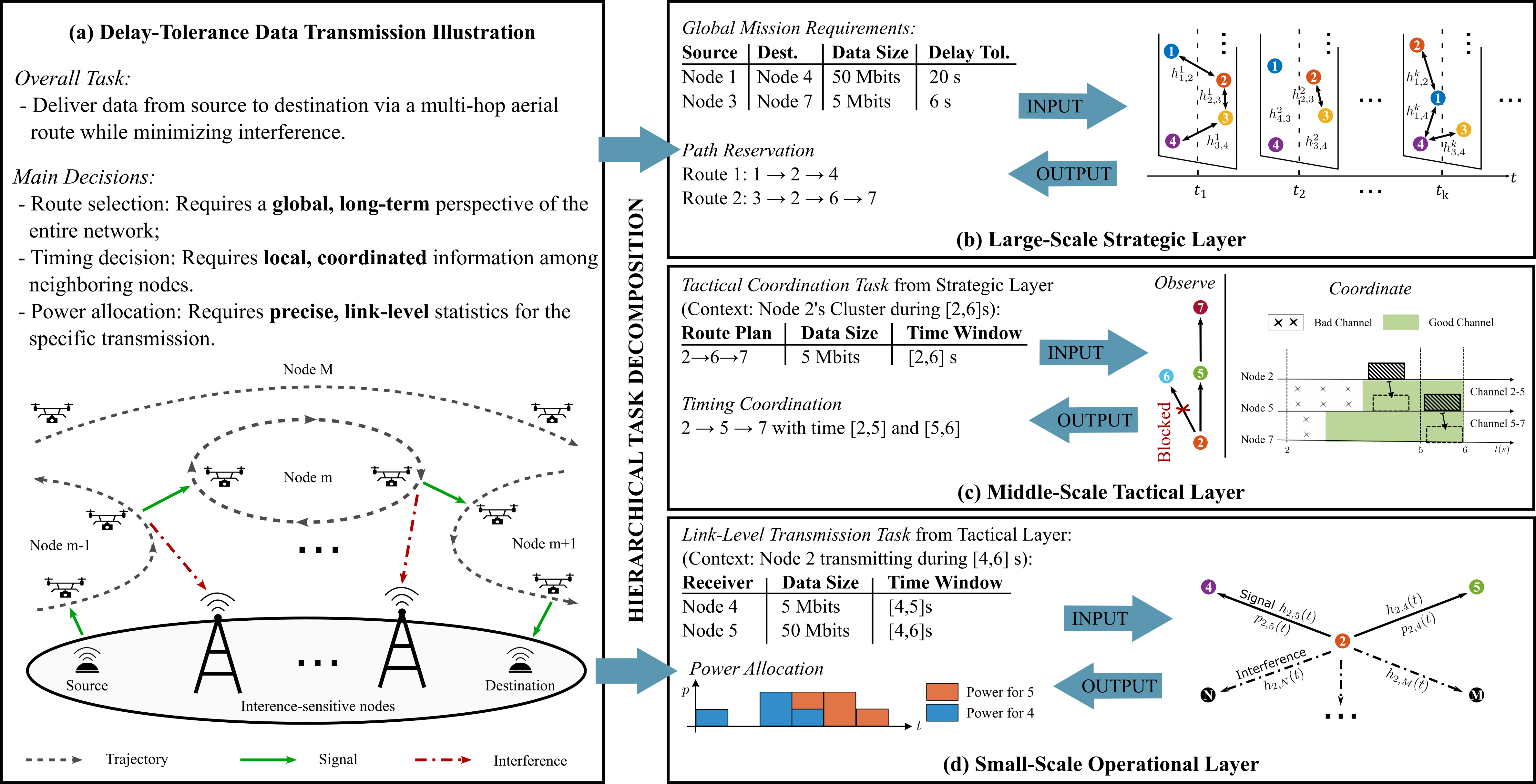}
\par\end{centering}
\caption{\label{fig:framework}The layered predictive communication framework, illustrating the hierarchical task decomposition for a proactive communication task. (a) The overall objective is to deliver data from a source to a destination via a multi-hop aerial route while minimizing interference. The main decisions, route selection, timing, and power allocation, require information at different scales. This complex problem is decomposed across three layers: (b) The Large-Scale Strategic Layer uses global, long-term information to transform mission requirements into a reserved end-to-end path. (c) This path is passed as a directive to the Middle-Scale Tactical Layer, where a local cluster uses more precise, local-area predictions to coordinate the timing of each hop, potentially adjusting the route to bypass a blockage. (d) Finally, the Small-Scale Operational Layer receives a specific transmission schedule. It uses link-level statistics to design a predictive power allocation policy that optimizes physical-layer efficiency while managing interference.}
\end{figure*}

\subsection{Accuracy vs. Range Trade-Off}

The interplay between the physical capabilities of the network echelons and the fundamental sources of uncertainty gives rise to a distinct information hierarchy, defined by a crucial trade-off between predictive accuracy and range. 

{\em The central echelon is characterized by the maximum predictive range but the lowest accuracy.} At the apex, the central echelon, typically a central control authority, retains every pre-filed 4D trajectory and a city-scale radio map, offering a network-wide, long-horizon predictive view. Its range is a direct consequence of its physical role as the global repository for all pre-planned mission trajectories, which grants it an unparalleled, long-horizon view of the entire network's planned evolution. However, this reliance on a global, planned-level abstraction is the source of its low predictive fidelity. Its model must operate on planned rather than real-time trajectories, forcing it to treat on-the-fly deviations as a statistical variance. Furthermore, its large-area environmental model is subject to information staleness and inaccuracies from periodic, aggregated reporting. Consequently, the central forecast must account for the full spectrum of these uncertainties, yielding a high-variance statistical model.

{\em The local echelon, in contrast, exhibits intermediate range and significantly improved accuracy.} Serving as an intermediary, the local echelon, often a ground base station or edge server, aggregates high-rate feedback, such as the real-time positions of neighboring nodes, from a regional cluster of aircraft, refining the central forecast with low-latency, high-fidelity local data. Its range is confined to a local subgraph, such as the aircraft within a base station's coverage area. Its physical capability as a regional information aggregator allows it to gather real-time positional data from its local cluster, thereby eliminating the trajectory deviation uncertainty that plagues the central tier. While its predictive range is limited to its regional sub-graph, the resulting lower-variance forecast is more reliable.

{\em Finally, the individual echelon has the most limited range but the highest accuracy.} At the base is the individual echelon, consisting of each aircraft itself; equipped with onboard sensors, it performs instantaneous channel measurements and precise self-localization. Its predictive scope is restricted to its egocentric, one-hop neighborhood. However, its onboard capability to perform direct, instantaneous channel measurements provides a ground-truth anchor for its forecasts. This allows it to temporarily eliminate all sources of uncertainty, yielding a predictive model with a unique, near-zero initial variance.

Therefore, this clear hierarchy of capabilities logically dictates a corresponding function hierarchy. The central tier's broad but coarse foresight is well-suited for strategic, network-wide planning. The local tier's balance of regional scope and tactical-grade fidelity is ideal for multi-node tactical coordination. Lastly, the individual tier's highly precise but myopic foresight perfectly matches the link-level policy optimization task. This functional decomposition, summarized in Table~\ref{tab:h_nature}, forms the guiding principle for our proactive control framework.

\begin{figure*}
\begin{centering}
\includegraphics[width=1\textwidth]{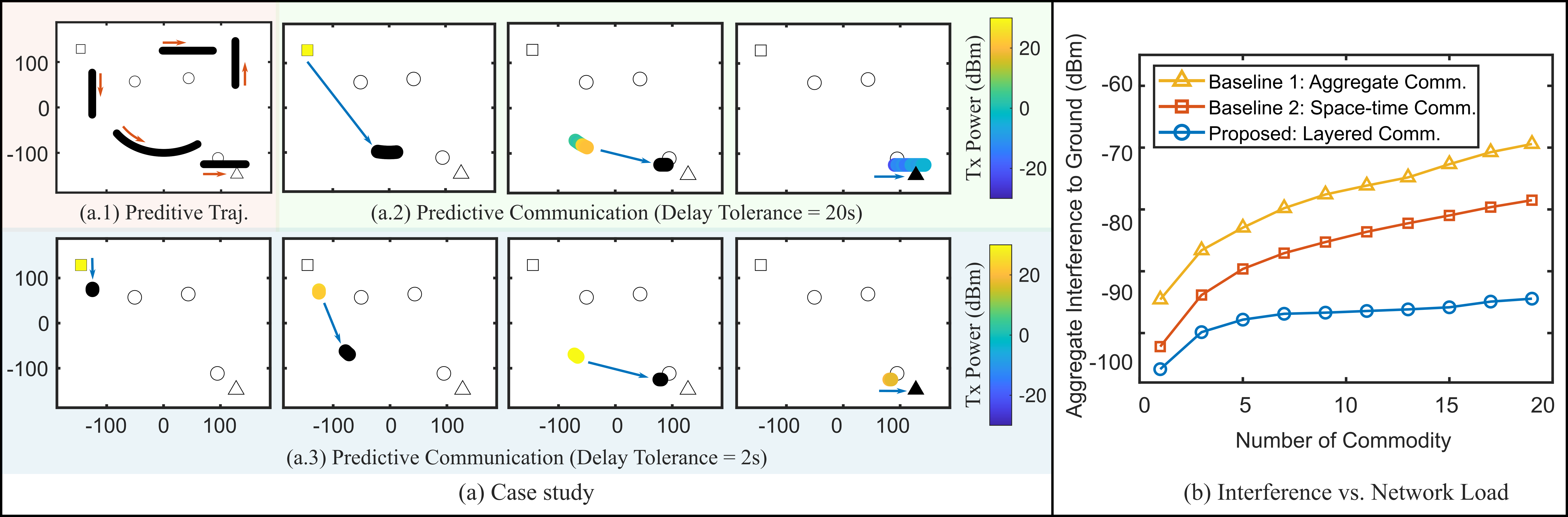}
\par\end{centering}
\caption{\label{fig:routing_illustration}{Interference-aware predictive communications. (a) Case study. (a.1) Considered scenario with predictive aircraft mission trajectories, where black dots denote aircraft positions and red arrows indicate flight directions, over a ground layout comprising a source (square), a destination (triangle), and interference-sensitive nodes (circles). (a.2) Planned route and associated transmit power (color bar) for a delay-tolerant task (delay tolerance 20 s). (a.3) Planned route and transmit power for a delay-sensitive task (delay tolerance 2 s). (b) Aggregate interference power of the proposed method versus two baselines under increasing network load (number of commodities).}}
\end{figure*}

\section{Layered Predictive Communication Framework}

The core principle of this framework is hierarchical task decomposition, as conceptually illustrated in Fig.~\ref{fig:framework}. The predictive cross-layer objective of delivering data with minimal interference (Fig.~\ref{fig:framework}(a)) is broken down into a cascade of distinct, tractable sub-problems. This is achieved by assigning different functional responsibilities to three layers, each operating at a different scale and leveraging the predictive information best suited to its task.
{Within this hierarchy, channel uncertainty caused by trajectory deviations, environmental modeling errors, and small-scale fading is captured through a statistical characterization of the problem, {\em e.g.}, expectation- or outage-based objectives and constraints.}

\subsection{The Large-Scale Strategic Layer}

The strategic layer performs long-term, network-wide path reservation. Operating at the apex of the hierarchy, it leverages a global, long-horizon predictive view to optimize end-to-end routes, a task impossible for lower echelons with only localized information. While its predictive model is statistically coarse, it is uniquely capable of establishing robust, high-level directives for the entire network. For example, as illustrated in Fig.~\ref{fig:framework}(b), the strategic layer translates global mission requirements into reserved paths using the global spatio-temporal channel graph.
{By incorporating threat intelligence into the planning process, the strategic layer can steer entire data flows around known jamming zones or regions with high interception risk, explicitly trading off path length for improved security.}

\subsection{The Middle-Scale Tactical Layer}

{The tactical layer adapts the strategic plan to dynamic local realities using more accurate, mid-term predictions. Its key functions are multi-node coordination, reactive re-routing, and proactive re-timing.}
A canonical example is illustrated in Fig.~\ref{fig:framework}(c). The strategic layer may have issued a route directive of $2\to 6 \to 7$. However, the nodes in the local cluster observe that the direct link $2 \to 6$ is {physically blocked by a building or virtually blocked by a suspected jamming attack or a newly imposed "no-fly" zone.} 
Therefore, the tactical layer's first action is to perform a local re-routing, adjusting the path to $2\to 5 \to 7$ to bypass the obstruction. Having established a viable new path, it then proceeds to coordinate the transmission timing. Using their superior local predictive models, nodes 2 and 5 can identify and schedule their respective transmissions to coincide with periods of high channel gain and low security exposure. This dual capability of reactive re-routing and proactive re-timing allows the system to enhance efficiency and resilience in a way the coarse global model could not foresee.

\subsection{The Small-Scale Operational Layer}

At the base of the hierarchy, the operational layer tackles short-term, link-level optimization. {Using the most accurate, measurement-anchored statistics, each aircraft pre-computes an optimal transmission policy.} In Fig.~\ref{fig:framework}(d), the node leverages high-fidelity predictions of its desired link and the air-to-ground interference link to derive an interference-aware power control rule ahead of transmission. {This same predictive mechanism is vital for enhancing confidentiality. By also predicting the channel to a potential eavesdropper, the aircraft can design a low probability of intercept power policy, dynamically adjusting its transmit power to meet its own communication needs while remaining below the eavesdropper's detection threshold. This enables energy-efficient, and secure resource allocation.}

{This layered decomposition is the key to a strategically robust yet tactically agile system. The strategic layer uses long-term predictive information to set global routing goals and constraints. In contrast, the tactical and operational layers refine these plans with progressively shorter prediction-execution gaps using fresher channel observations. In this way, long-horizon decisions are continuously corrected at lower layers, so the network remains robust to trajectory and environmental uncertainties while reacting quickly to real-world conditions.}

\section{Case Study {and Security Extensions}}

This section validates our layered predictive framework through a case study on interference mitigation \cite{LiChe:J26}, with key results illustrated in Fig.~\ref{fig:routing_illustration}. First, the large-scale layer demonstrates strategic adaptability, exploiting aircraft mobility for delay-tolerant tasks (Fig.~\ref{fig:routing_illustration}(a.2)) while using multi-hop relays for delay-sensitive tasks (Fig.~\ref{fig:routing_illustration}(a.3)). The middle-scale layer then performs tactical timing coordination based on channel dynamics, scheduling a high-power, rapid transmission for the degrading first two hops in Fig.~\ref{fig:routing_illustration}(a.2) while using a low-power, longer transmission for the improving last hop. The small-scale layer ensures fine-grained interference control, strategically reducing transmit power when aircraft operate near sensitive ground nodes. 
{Compared with the classical aggregate-based baseline (routing on a static graph with locally coordinated handovers) and the space-time-based baseline (routing on a space-time graph with fixed, equally spaced handover times),} the proposed layered strategy consistently achieves a larger reduction in aggregate interference power, as validated in Fig.~\ref{fig:routing_illustration}(b), with this advantage becoming more pronounced as network load increases. {Note that goal-oriented network formulations reduce the computational burden and keep centralized implementation tractable.}

Although this case study focuses on interference-aware communications, the same layered predictive framework is also well-suited for security-oriented designs in low-altitude networks. By appropriately augmenting the optimization objectives and constraints with security-related metrics, predictive routing, handover timing, and power control can be tailored to minimize exposure to suspected jamming or high-risk regions in space and time; radio-map-assisted channel prediction can serve as a prior for anomaly detection in positioning and control signals ({\em e.g.}, GPS spoofing or link hijacking); and the hierarchical timing structure naturally supports geofencing- and time-aware scheduling policies that avoid transmitting sensitive data over vulnerable links or areas, without altering the core architecture of the framework.

\section{Challenges and Opportunities}

There are multiple future directions for predictive communications in low-altitude networks worthy of further investigation, as discussed below.

\subsection{Dynamic Environmental Awareness}

Our framework demonstrates the power of leveraging static environmental models. A critical next step is to enable these environmental representations to dynamically update in real time, transforming them into "living" models of the world. This introduces key challenges in efficient data management. Semantic communication techniques are needed on the uplink to allow aircraft to intelligently identify and report only the most salient environmental changes (e.g., terrain and weather), rather than raw \ac{csi} measurements. On the downlink, efficient dissemination of model updates is crucial, potentially by partitioning the model into standard (global) and private (local) components to reduce redundancy. This would create a system with continuously evolving environmental awareness, making its predictions even more accurate and adaptive.

\subsection{Predictive Coverage}

The predictive paradigm can be extended beyond routing and resource allocation to optimize network coverage itself proactively. This can be approached from two angles. First, predictive interference maps can inform the long-term planning and adjustment of terrestrial base station parameters (e.g., antenna tilts) to serve heavy-traffic air corridors better. Second, the framework can guide the dynamic deployment and trajectory planning of aerial base stations, positioning them not where coverage is currently weak, but where it is predicted to be needed, proactively filling anticipated gaps in terrestrial service.

\subsection{Predictive Integrated Sensing and Communication (ISAC)}

Integrating sensing into the predictive framework offers a powerful way to mitigate the massive data redundancy inherent in multi-aircraft ISAC. The key is to co-schedule sensing and communication tasks by leveraging foresight in both temporal and spatial dimensions. Temporally, the framework can predict future windows of high-quality sensing channels (e.g., optimal target angle) and communication channels, enabling a sparse schedule of high-value sensing actions and opportunistic, low-power transmissions for a single aircraft. Spatially, it can predict future overlaps in the sensing fields of view over multiple aircraft. This foresight enables proactive coordination, such as assigning specific sensing sectors to each aircraft or selecting a single node to measure. In contrast, others relay the data, thus eliminating redundant sensing and transmission.

\subsection{{Predictive Security and Resilience}}
{
While this paper introduces the potential for predictive security, it opens a rich field for future work. Research is needed to develop sophisticated threat models that can be integrated into the predictive objectives, moving beyond simple high-risk zones to probabilistic models of attacker behavior. Furthermore, leveraging the framework for distributed, multi-agent security games, where aircraft learn to cooperatively detect and isolate threats in a decentralized manner, represents a vital direction for enhancing network resilience.}

\section{Conclusion}

This paper proposes a predictive communication framework for low-altitude networks, which leverages hierarchical predictive information to derive robust strategies for routing, coordination, and policy design. To spur further innovation, we outline several key future research directions that will support and promote the continued development of this framework.

\bibliographystyle{IEEEtran}
\bibliography{IEEEabrv, StringDefinitions, JCgroup, JCgroup-bw}

@STRING{IEEE_J_VT         = "{IEEE} Trans. Veh. Technol."}

@STRING{IEEE_J_SP         = "{IEEE} Trans. Signal Process."}

@STRING{IEEE_J_COM        = "{IEEE} Trans. Commun."}

@STRING{IEEE_J_WCOM       = "{IEEE} Trans. Wireless Commun."}

@STRING{IEEE_M_COM        = "{IEEE} Commun. Mag."}

@STRING{IEEE_O_CSTO        = "{IEEE} Commun. Surveys Tuts."}

@article{LiuXuWanChe:J20,
  author={Liu, Dianxiong and Xu, Yuhua and Wang, Jinlong and Chen, Jin and Yao, Kailing and Wu, Qihui and Anpalagan, Alagan},
  journal=IEEE_M_COM, 
  title={Opportunistic {UAV} Utilization in Wireless Networks: Motivations, Applications, and Challenges}, 
  year={2020},
  volume={58},
  number={5},
  pages={62-68},
  doi={10.1109/MCOM.001.1900687}
}

@ARTICLE{LiChe:J24,
  author={Li, Bowen and Chen, Junting},
  journal=IEEE_J_WCOM, 
  title={Radio Map-Assisted Approach for Interference-Aware Predictive {UAV} Communications}, 
  year={2024},
  volume={23},
  number={11},
  pages={16725-16741}}

@ARTICLE{SunChe:J24,
  author={Sun, Hao and Chen, Junting},
  journal=IEEE_J_SP, 
  title={Integrated Interpolation and Block-Term Tensor Decomposition for Spectrum Map Construction}, 
  year={2024},
  volume={72},
  number={},
  pages={3896-3911}
}

@ARTICLE{LiChe:J24b,
  author={Li, Bowen and Chen, Junting},
  journal=IEEE_J_COM, 
  title={Large Timescale Optimization for Communications Over Aerial Ad Hoc Networks With Predetermined Trajectories}, 
  year={2024},
  volume={72},
  number={10},
  pages={6371-6385}}

@ARTICLE{HanXuZhaWan:J23,
  author={Han, Zhenzhen and Xu, Chuan and Zhao, Guofeng and Wang, Shanshan and Cheng, Kefei and Yu, Shui},
  journal=IEEE_J_VT, 
  title={Time-Varying Topology Model for Dynamic Routing in {LEO} Satellite Constellation Networks}, 
  year={2023},
  volume={72},
  number={3},
  pages={3440-3454},
  doi={10.1109/TVT.2022.3217952}
}

@ARTICLE{VaeLinZha:J24,
  author={Vaezi, Mojtaba and Lin, Xingqin and Zhang, Hongliang and Saad, Walid and Poor, H. Vincent},
  journal=IEEE_M_COM, 
  title={Deep Reinforcement Learning for Interference Management in {UAV}-Based {3D} Networks: Potentials and Challenges}, 
  year={2024},
  volume={62},
  number={2},
  pages={134-140},
  doi={10.1109/MCOM.001.2200973}}

@ARTICLE{SinKumMajSat:J25,
  author={Singh, Shivani and Kumar, Sushant and Majhi, Sudhan and Satija, Udit and Yuen, Chau},
  journal=IEEE_O_CSTO, 
  title={Blind Carrier Frequency Offset Estimation Techniques for Next-Generation Multicarrier Communication Systems: Challenges, Comparative Analysis, and Future Prospects}, 
  year={2025},
  volume={27},
  number={1},
  pages={1-36}
  }

@ARTICLE{JiaLiZhuLi:M25,
  author={Jiang, Yihang and Li, Xiaoyang and Zhu, Guangxu and Li, Hang and Deng, Jing and Han, Kaifeng and Shen, Chao and Shi, Qingjiang and Zhang, Rui},
  journal=IEEE_M_COM, 
  title={Integrated Sensing and Communication for Low Altitude Economy: Opportunities and Challenges}, 
  year={2025},
  volume={},
  number={},
  pages={1-7},
  doi={10.1109/MCOM.001.2400685}
}

@ARTICLE{SonLinWanSun:M25,
  author={Song, Mengshu and Lin, Yijing and Wang, Jiacheng and Sun, Geng and Dong, Chen and Ma, Nan and Niyato, Dusit and Zhang, Ping},
  journal=IEEE_M_COM, 
  title={Trustworthy Intelligent Networks for Low-Altitude Economy}, 
  year={2025},
  volume={63},
  number={7},
  pages={72-79},
  doi={10.1109/MCOM.001.2400692}
}

@ARTICLE{HanJiaYu:J25,
  author={Han, Han and Jiang, Tao and Yu, Wei},
  journal=IEEE_J_WCOM, 
  title={Active Sensing for Multiuser Beam Tracking With Reconfigurable Intelligent Surface}, 
  year={2025},
  volume={24},
  number={1},
  pages={540-554},
  doi={10.1109/TWC.2024.3496393}
  }

@ARTICLE{WanZhaNieYu:M25,
  author={Wang, Heng and Zhang, Jianhua and Nie, Gaofeng and Yu, Li and Yuan, Zhiqiang and Li, Tongjie and Wang, Jialin and Liu, Guangyi},
  journal=IEEE_M_COM, 
  title={Digital Twin Channel for 6{G}: Concepts, Architectures and Potential Applications}, 
  year={2025},
  volume={63},
  number={3},
  pages={24-30},
  doi={10.1109/MCOM.001.2400213}
  }

@ARTICLE{HeAiGuaWan:J19,
  author={He, Danping and Ai, Bo and Guan, Ke and Wang, Longhe and Zhong, Zhangdui and K\"{u}rner, Thomas},
  journal=IEEE_O_CSTO, 
  title={The Design and Applications of High-Performance Ray-Tracing Simulation Platform for {5G} and Beyond Wireless Communications: A Tutorial}, 
  year={2019},
  volume={21},
  number={1},
  pages={10-27},
  doi={10.1109/COMST.2018.2865724}
}

@ARTICLE{SonZenYanRen:J25,
  author={Song, Yuxuan and Zeng, Yong and Yang, Yuhang and Ren, Zixiang and Cheng, Gaoyuan and Xu, Xiaoli and Xu, Jie and Jin, Shi and Zhang, Rui},
  journal=IEEE_M_COM, 
  title={An Overview of Cellular {ISAC} for Low-Altitude {UAV}: New Opportunities and Challenges}, 
  year={2025},
  volume={},
  number={},
  pages={1-8},
  doi={10.1109/MCOM.002.2400742}}

@INPROCEEDINGS{NtoTomRui:C24,
  author={Ntontin, K. and Tomaszewski, L. and Ruiz-de-Azua, J. A. and C\'{a}rdenas, A. and Centelles, R. Pueyo and Lin, C.-K. and Mesodiakaki, A. and Antonopoulos, A. and Pappas, N. and Fiore, M. and Aguilar, S. and Watts, S. and Harris, P. and Santiago, A. R. and Lazarakis, F. and Calisti, M. and Chatzinotas, S.},
  booktitle=WCNC, 
  title={ETHER: A {6G} Architectural Framework for {3D} Multi-Layered Networks}, 
  year={2024},
  volume={},
  number={},
  pages={1-6},
  doi={10.1109/WCNC57260.2024.10570731}
}

@ARTICLE{LiChe:J26,
  author={Li, Bowen and Chen, Junting},
  journal=IEEE_J_WCOM, 
  title={Radio Map-Assisted Routing and Predictive Resource Allocation Over Dynamic Low-Altitude Networks}, 
  year={2026},
  volume={25},
  number={},
  pages={9955-9970}}

@STRING{IEEE_J_WCOM = {IEEE Trans. on Wireless Commun.}}

@STRING{IEEE_J_COM = {IEEE Trans. on Commun.}}

@STRING{AI = {Artif. Intell.}}

@STRING{WCNC = {Proc. IEEE Wireless Commun. and Networking Conf.}}

\section*{Biographies}

\textbf{Junting Chen} (S'11--M'16) received the Ph.D.\ degree in Electronic and Computer Engineering from the Hong Kong University of Science and Technology (HKUST), Hong Kong SAR China, in 2015, and the B.Sc.\ degree in Electronic Engineering from Nanjing University, Nanjing, China, in 2009. From 2014--2015, he was a visiting student with the Wireless Information and Network Sciences Laboratory at MIT, Cambridge, MA, USA.  

He is an Assistant Professor with the School of Science and Engineering and the Shenzhen Future Network of Intelligence Institute (FNii--Shenzhen) at The Chinese University of Hong Kong, Shenzhen (CUHK--Shenzhen), Guangdong, China. Prior to joining CUHK--Shenzhen, he was a Postdoctoral Research Associate with the Ming Hsieh Department of Electrical Engineering, University of Southern California (USC), Los Angeles, CA, USA, from 2016--2018, and with the Communication Systems Department of EURECOM, Sophia--Antipolis, France, from 2015--2016. His research interests include channel estimation, MIMO beamforming, machine learning, and optimization for wireless communications and localization. His current research focuses on radio map sensing, construction, and application for wireless communications. Dr. Chen was a recipient of the HKTIIT Post-Graduate Excellence Scholarships in 2012. He was nominated as the Exemplary Reviewer of {\scshape IEEE Wireless Communications Letters} in 2018. His paper received the Charles Kao Best Paper Award from WOCC 2022.

\textbf{Bowen Li} (Member, IEEE) received the Ph.D.\ degree in Computer and Information Engineering from The Chinese University of Hong Kong, Shenzhen (CUHK--Shenzhen), China, in 2025. He is  currently a Postdoctoral Research Associate with Department of Computer and Information Science, Link\"{o}ping University. His research interests primarily include predictive communications, semantic communications, and signal processing for low-altitude networks.

\textbf{Hao Sun} received the B.S. degree from the University of Electronic Science and Technology of China (UESTC), Chengdu, China, in 2018. He received the Ph.D. degree in Computer and Information Engineering from The Chinese University of Hong Kong, Shenzhen (CUHK--Shenzhen), Guangdong, China, in 2025. He is currently a Postdoctoral Fellow with the Department of Electrical Engineering, City University of Hong Kong, Hong Kong. His research interests include matrix completion, tensor decomposition, and generative AI with applications with applications in wireless communication and signal processing.

\textbf{Shuguang Cui} (Fellow, IEEE) received the Ph.D. degree from Stanford University in 2005. He is currently a X. Q. Deng Presidential Chair Professor at The Chinese University of Hong Kong, Shenzhen, China. His current research interests include data driven large-scale information analysis and system design. He was selected as the Thomson Reuters Highly Cited Researcher and listed in the Worlds' Most Influential Scientific Minds by ScienceWatch in 2014. He was a recipient of the IEEE SP Society 2012 and ComSoc 2023 Marconi Best Paper Awards.
He is a Member of Both Royal Society of Canada and Canadian Academy of Engineering.

\textbf{Nikolaos Pappas} (Senior Member, IEEE) is currently an Associate Professor with the Department of Computer and Information Science, Link\"{o}ping University, Link\"{o}ping, Sweden. His research interests include the field of wireless communication networks, with an emphasis on semantics-aware communications, energy harvesting networks, age of
information, and stochastic geometry. He has served as the Symposium Co-Chair for the IEEE International Conference on Communications in 2022. He
was the General Chair for the 23rd International Symposium on Modeling and Optimization in Mobile, Ad hoc, and Wireless Networks (WiOpt 2025). He is an Area Editor of IEEE OPEN JOURNAL OF THE COMMUNICATIONS SOCIETY and an Expert Editor of invited papers of IEEE COMMUNICATIONS LETTERS. He is an Associate Editor of four IEEE TRANSACTIONS and journals.
\end{document}